\documentclass[aps, prb, citeautoscript, amsmath, amssymb, reprint]{revtex4-1}
\usepackage{graphicx}
\usepackage{color}
\usepackage{cases,array}
\usepackage{bm}

\usepackage[english]{babel}
\usepackage{float}

\usepackage{hyperref}

\usepackage{commath}

\begin{document}

\title{Driven energy transfer between coupled modes in spin-torque oscillators}

\author{R. Lebrun}
\author{J. Grollier}
\author{P. Bortolotti}
\author{V. Cros}
\affiliation{Unit\'e Mixte de Physique CNRS, Thales, Univ. Paris-Sud, Universit\'e Paris-Saclay, Palaiseau, France}

\author{A. Hamadeh}
\author{X. de Milly} 
\author{Y. Li}
\author{G. de Loubens}
\affiliation{Service de Physique de l'\'Etat
Condens\'e, CEA, CNRS, Universit\'e Paris-Saclay, Gif-sur-Yvette, France}

\author{O. Klein}
\affiliation{SPINTEC, UMR CEA/CNRS/UJF-Grenoble 1/Grenoble-INP, Grenoble, France}

\author{S. Tsunegi}
\author{H. Kubota}
\author{K. Yakushiji}
\author{A. Fukushima}
\author{S. Yuasa}
\affiliation{National Institute of Advanced Industrial Science and Technology (AIST), Spintronics Research Center, Tsukuba, Japan}

\altaffiliation{corresponding author : vincent.cros@thalesgroup.com}

\begin{abstract}
The mutual interaction between the different eigenmodes of a spin-torque oscillator can lead to a large variety of physical mechanisms from mode hopping to multi-mode generation, that usually reduce their performances as radio-frequency devices. To tackle this issue for the future applications, we investigate the properties of a model spin-torque oscillator that is composed of two coupled vortices with one vortex in each of the two magnetic layers of the oscillator. In such double-vortex system, the remarkable properties of energy transfer between the coupled modes, one being excited by spin transfer torque while the second one being damped, result into an alteration of the damping parameters. As a consequence, the oscillator nonlinear behavior is concomitantly drastically impacted. This efficient coupling mechanism, driven mainly by the dynamic dipolar field generated by the spin transfer torque induced motion of the vortices, gives rise to an unexpected dynamical regime of self-resonance excitation.  These results show that mode coupling can be leveraged for controlling the synchronization process as well as the frequency tunability of spin-torque oscillators.
\end{abstract}

\maketitle


For the last decade, spin-torque oscillators (STOs) \cite{kiselev_2003} have attracted a large interest as they have been considered not only as model systems to study nonlinear dynamics at the nanoscale but also as promising candidates for a new generation of radiofrequency devices\cite{locatelli_2014-1}. These highly nonlinear magnetic oscillators, compatible with CMOS technology, possess a large frequency tunability. However the control of their microwave features, and especially of their spectral coherence, is far from being fully understood \cite{romera_2015, gusakova_2009, hamadeh_2014-1, muduli_2012} and from reaching the requirements for real microwave applications. As for any non-magnetic oscillators \cite{okamoto_2013, yariv_1973}, the poor control of their microwave specifications can originate from an unwanted transfer of energy between the different eigenmodes of the system.

In the past few years, the presence of multi-mode generation in STOs, with either mode hopping or mode coexistence, has been reported \cite{iacocca_2014, heinonen_2013, lee_2012, muduli_2012, gusakova_2009, gusakova_2011, iacocca_2014, slobodianiuk_2014, hamadeh_2014-1}. These complex spin transfer dynamics result from a transfer of energy between different magnetic modes of the system that is particularly efficient when the mode frequencies cross each other \cite{pigeau_2012, lequeux_2015, jung_2011, sugimoto_2011}. This mode coupling generally comes at the cost of a degradation of the STOs' performances as highlighted by numerous studies \cite{collet_2015, hamadeh_2014-1, gusakova_2009, gusakova_2011, iacocca_2014, romera_2015}. However, some other studies have highlighted that mode hybridization could lead to a reduction of the oscillator nonlinear parameters, and thereby enhance the oscillator spectral coherence \cite{hamadeh_2014-1, lebrun_2015-1}. This apparent discrepancy reveals that the physical origins of the coupling and its consequences on the high frequency dynamics of STOs remain a crucial issue to be elucidated.

For STOs, the interaction between several modes can have several origins and be for example either mediated by spin-waves \cite{mancoff_2005}, exchange \cite{iacocca_2014, ruotolo_2009}, dipolar interaction \cite{pigeau_2012, locatelli_2015, hamadeh_2014-1, keatley_2013, jung_2011, sugimoto_2011} or electrically through the spin transfer torque (STT) induced emitted rf current\cite{grollier_2006, lebrun_2016}. As the spin transfer torque selectively excites the different magnetic modes depending on their symmetry, some of the eigenmodes are effectively enhanced by STT \cite{iacocca_2015}, while some others are damped. Combined with the difficulty to simultaneously access the properties of the both eigenmodes excited and damped by STT \cite{naletov_2011}, it has largely limited the understanding of the mode coupling mechanisms \cite{iacocca_2014, iacocca_2015, locatelli_2014, gusakova_2011, hamadeh_2014-1}. Identifying means to efficiently control the coupling and the properties of the different magnetic modes is thus a necessary step to further understand and control the nonlinear magnetization dynamics of STOs. This is also crucial for the development of novel functionalities, often related to the dynamics of a large number of spin torque oscillators \cite{locatelli_2014-1}.
 
\begin{figure*}
\centering
\includegraphics[scale=0.32]{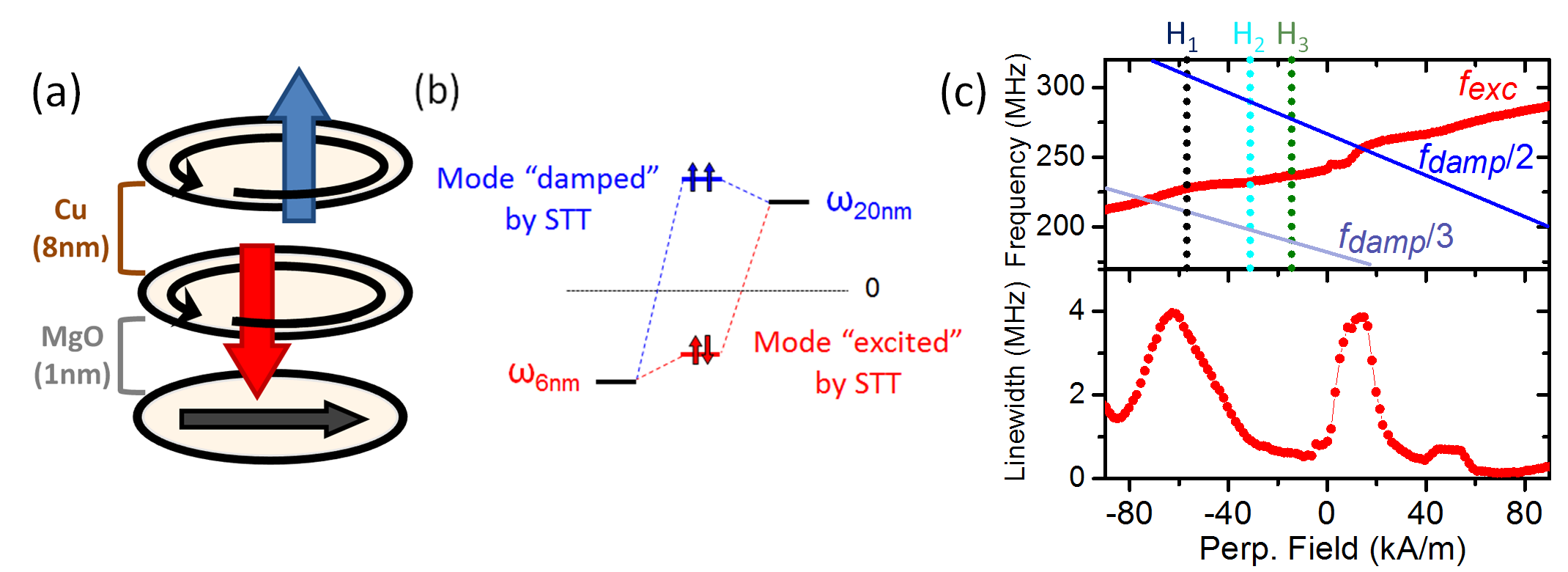}
\caption{(a) Schematic of an hybrid magnetic tunnel junction: a Cu based spin-valve system with the two vortex Py layers (20 nm at the top, 6 nm at the bottom) above a 1 nm MgO barrier and a CoFeB based synthetic antiferromagnet. (b) Sketch of the hybridized gyrotropic modes of the double vortex system. The thin (thick) vortex layer and its associated coupled mode are excited (damped) by STT. (c) Evolution of the frequency $f_{exc}$, the linewidth and the power of the excited mode as a function of the external perpendicular field for $I_{dc}=+11$ mA. The frequency of the mode that is damped by STT is called $f_{damp}$ ($H_1$, $H_2$, $H_3$ correspond to the fields that we study by time domain measurements in Fig. \ref{fig:timedomain_crossing}).}
\label{fig:fig1}
\end{figure*}

In this article, our main objective is to investigate experimentally the impact of mode hybridization through the study of the dynamical properties of a double vortex based STO that has two weakly coupled gyrotropic eigenmodes. These specific STOs also present the advantage to have a sufficiently large emitted power (because of the large tunnel magnetoresistance ratio) to directly correlate through time domain measurements the coupling between modes and the oscillator’s nonlinear behavior. Indeed energy transfer between the two coupled modes leads to a large modification of both the magnetization damping and the amplitude relaxation rate of the mode that are excited by spin transfer torque. These results highlight some of the limitations of the single mode approach for describing the behavior of a STO. In fact, it is not sufficient to simply consider the magnetic layer excited by STT in order to model properly the dynamics of the oscillator. In our STO devices (double spin-valve above a magnetic tunnel junction), we are able to simultaneously study the dynamics of the two inter-layer coupled vortex modes, while one mode is damped and the other one is excited by STT. We identify the main mechanism that couples the dynamics of the two modes and drives the transfer of energy between them. Through a fine control of the mode frequencies, we also report a new phenomenon that we call “self-resonant excitation”, for which the dynamics of the damped mode is driven into a forced oscillating regime through its interaction with the excited mode. We believe that such self-resonant process is of high interest as it opens a new approach to develop multiband STOs.

\section{Tunable nonlinear behavior driven by mode coupling}

The magnetic system consists of two magnetic vortex layers separated by a Cu layer on top of a MgO barrier and a synthetic antiferromagnet (SAF) NiFe(20 nm)/Cu(8 nm)/NiFe(6 nm)/MgO(1 nm)/SAF. The two magnetic vortex layers are weakly coupled (see Fig. \ref{fig:fig1}) through dipolar interaction between the vortex cores and vortex bodies. This leads to an oscillator system in which the two gyrotropic modes of each vortex layer are weakly hybridized. Detailed description of this double-vortex system and its basic dynamic properties can be found elsewhere \cite{lebrun_2014}.
 
In a system containing two vortices, several static magnetic configurations are accessible depending on the respective core polarities and vortex chiralities. In this study, we focus on the case of two vortices with parallel chiralities (defined by the current sign through the Oersted field) and antiparallel core polarities for which sustained STT oscillations can be obtained at zero or low perpendicular magnetic field \cite{lebrun_2014, locatelli_2011}. In the presence of electrical current, each vortex layer plays the role of a spin polarizer for the other vortex layer. For the current sign that we have chosen, the STT induces oscillations corresponding to the coupled mode dominated by the thin layer (schematized in red in Fig. \ref{fig:fig1}.b). The other mode associated with the coupled mode dominated by the thick vortex layer (schematized in blue in Fig. \ref{fig:fig1}.b) is damped by STT \cite{lebrun_2014}.

The magnetoresistive ratio associated with the magnetic tunnel junction is much larger than the one of the spin-valve part ($80 \%$ compared to $3 \%$). Consequently, the emitted power (of a few hundred nanowatts) of the STO is dominated by the vortex dynamics in the 6 nm NiFe layer adjacent to the MgO barrier layer (see Fig. \ref{fig:fig1}.a-b). Note that the spin-polarized current originating from the SAF layers have no STT contribution in the absence of out-of-plane component of the magnetization \cite{dussaux_2012}. Therefore, the tunnel junction here only serves as a sensitive detector of magnetization dynamics.

In Fig. \ref{fig:fig1}.c, we present the properties of the coupled mode that is excited by STT as a function of the perpendicular field $H_{perp}$. First we see that its frequency $f_{exc}$ increases, following an almost linear trend with $H_{perp}$.  This is expected as the dynamics of this coupled mode is indeed mainly driven by the bottom vortex layer whose core polarity is parallel to the positive applied field \cite{deloubens_2009}. The linewidth presents large variations with maxima in the regions for which the frequency evolution moves away from a linear behavior. Similar non-monotonic dependency of the spectral coherence vs $H_{perp}$ has been reported in different kinds of STOs and  linked to the interaction between the self-oscillating mode and other magnetic modes of the system \cite{hamadeh_2014-1, romera_2015, gusakova_2011}. In the case of a double vortex based STO, one of these other modes is the second coupled gyrotropic mode, damped by STT (see Fig. \ref{fig:fig1}.b). Note that in the present case, this second mode is dominated by the dynamics of the thick vortex layer. Moreover, given that the two vortices have opposite polarities, this frequency-field dispersion of this mode, represented as $f_{damp}$ on Fig. \ref{fig:fig1}.c, has a slope opposite to the one of the excited mode \cite{deloubens_2009}. Its frequency is determined theoretically \cite{lebrun_2014} and confirmed experimentally by the measurements presented in section B. The field regions of enhanced linewidths are precisely located at the harmonic crossings with this mode ($f_{damp}=nf_{exc}$ with $n$ integer as seen Fig. \ref{fig:fig1}.c) \cite{hamadeh_2014-1}. An important outcome of the present study is that we can correlate the nonlinear parameters of the auto-oscillating coupled gyrotropic mode with the coupling strength between this mode, excited by STT, and the mode damped by STT. The coupling strength is directly controlled by the perpendicular field applied on the system as it allows to bring the frequencies closer to each other. Then, for each of the coupling strentgh, we can perform systematic time domain measurements thanks to the high TMR ratio of our sensing junction, and, after Hilbert transformations, extract separately the power spectral densities of phase and amplitude noises \cite{quinsat_2012}.  

\begin{figure*}
\centering
\includegraphics[scale=0.47]{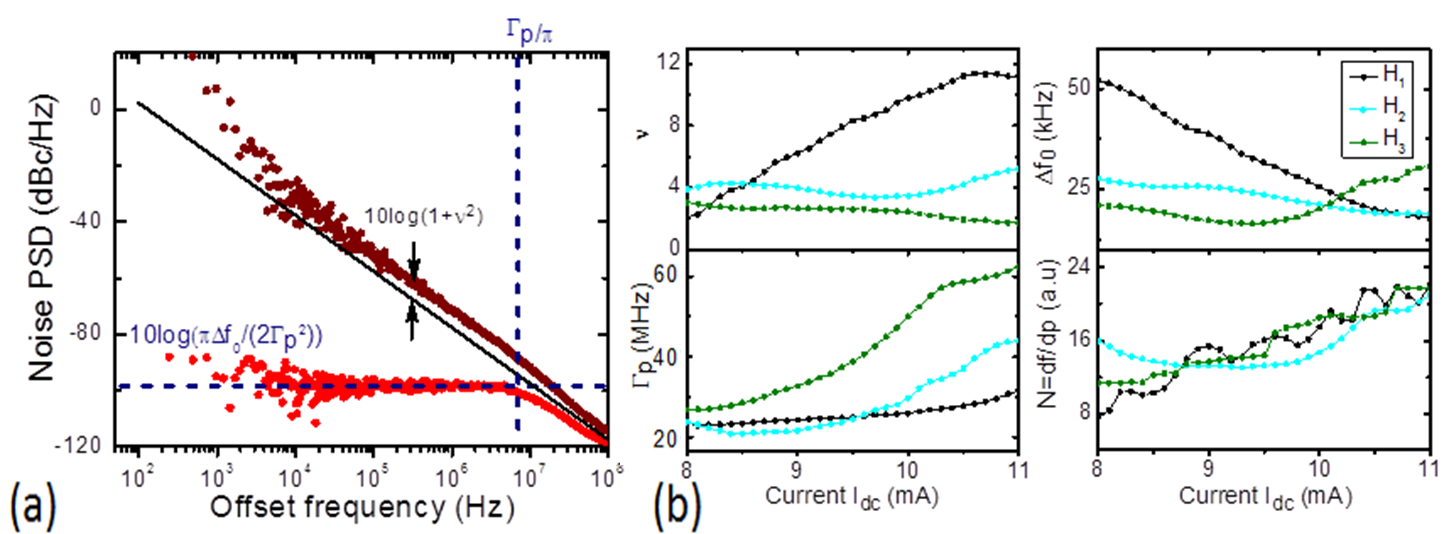}
\caption{(a) Noise power spectral density of the excited mode as a function of the offset frequency (for $I_{dc} = +8 $ mA and $H_{perp}=- 15 $ kA/m). The dark line corresponds to the intrinsic phase noise $\Delta f_0/\pi f^2$ (b) Evolution of the nonlinear parameters with $I_{dc}$ at three different fields: nonlinear dimensionless parameter $\nu$, the linear linewidth $\Delta f_0$, the amplitude relaxation rate $\Gamma_p$ and the nonlinear frequency shift $N$. The different fields are either close ($H_1=-55$ kA/m), far ($H_3=-15$ kA/m) from an harmonic crossing or between two crossings ($H_2=-30$ kA/m).}
\label{fig:timedomain_crossing}
\end{figure*}

Within the general model of auto-oscillators \cite{slavin_2009-1}, the dynamics of the oscillator can be described by the dynamics of its phase and of its normalized power $p=(\abs\rho/R)^2$ with $\rho$ the gyration radii and $R$ the dot radius. Its nonlinear behavior can then be characterized by a few key parameters. The first one is the nonlinear frequency shift $N$ that describes the frequency dependency with the amplitude of oscillations through the normalized power $p$. The second one is a relaxation rate that is called the amplitude relaxation rate $\Gamma_p$. It describes at which rate, the oscillator comes back to its limit cycle when it is perturbed by any external stimuli (including thermal ones). It is to be emphasized that this amplitude relaxation rate differs from the regular Gilbert-like relaxation as it includes both damping and STT contributions. This first two nonlinear parameters can be combined and allow to define a nonlinear dimensionless parameter $\nu = Np/\Gamma_p$ that represents the effective amplitude/phase coupling parameter. The third key parameter is the intrinsic linear linewidth $\Delta f_0$ due to thermal fluctuations. These three parameters $\nu$, $\Delta f_0$ and $\Gamma_p$ are extracted from the experimental diagrams of phase and amplitude noise \cite{grimaldi_2014-1} as shown in Fig. \ref{fig:timedomain_crossing}.a. The nonlinear frequency shift $N=df/dp$ can be also obtained from $df/dH_{perp}$ and $dp/dH_{perp}$ (as $df/dp= df/dH_{perp} \times{dH_{perp}/dp}$). 

Here after, we correlate the large variation of spectral linewidth with $H_{perp}$ shown in Fig. \ref{fig:fig1}.c with changes of the nonlinear parameters. In Fig. \ref{fig:timedomain_crossing}.b, we present the evolution of these  parameters with $I_{dc}$ determined experimentally for three different values of $H_{perp}$, i.e. three different coupling strengths between the two modes: (i) close ($H_1 = -55$ kA/m, black dots), (ii) far ($H_3=-15$ kA/m, green dots) or (iii) between ($H_2=- 30$ kA/m light blue dots)  one mode crossing as shown in Fig. \ref{fig:fig1}.c by vertical dotted lines.  

In the auto-oscillator model \cite{slavin_2009-1},the nonlinear parameters are expected to be proportional to $I_{dc}$. From the curves displayed in Fig. \ref{fig:timedomain_crossing}.b, we see that the experimental behaviors, even in the region of large mode interaction (that is for $H_1= -55$ kA/m), are not far from the predicted linear trend. The main deviation concerns the linear linewidth parameter $\Delta f_0$ that first decreases with $I_{dc}$ and then slightly increases at larger sur-criticalities ($I_{dc} > +10$ mA) in the region of weak interaction between the two modes, i.e. far from the crossing at $H_3 = -15$ kA/m. Similar behavior was observed in a single vortex STO (i.e. without the influence of interaction between several modes) by Grimaldi et al. \cite{grimaldi_2014-1} for large gyration radii. 
Hereafter, the striking feature is that the values of the oscillator parameters completely differ for the three different coupling strengths (fields) when the current increases. As shown in Fig. \ref{fig:timedomain_crossing}.b, for $I_{dc}=+11$ mA (similar to the case represented in Fig. \ref{fig:fig1}), the nonlinear frequency shifts $N$ are equal for the three magnetic fields and the intrinsic linewidths $\Delta f_0$ only slightly vary. On the contrary, the nonlinear dimensionless parameter $\nu$ and the amplitude relaxation rate $\Gamma_p$ vary strongly for the different fields. These observed variations of nonlinearities explain the large variation of linewidth as the expression of the STO linewidth is defined as $\Delta f=(1+\nu^2)\Delta f_0=(1+(Np_0/\Gamma_p)^2)\Delta f_0$ \cite{kim_2008}. In particular, the large $\nu$ and $\Gamma_p$ parameters observed for $H_1$, i.e. at harmonic crossing, are thus correlated with the large linewidth (around 4 MHz). 

In addition to these observations, we have measured larger threshold currents $I_{th}$ at the harmonic crossings ($I_{th}=+8$ mA at $-55$ kA/m against $I_{th}=+7$ mA at $-15$ kA/m). These lower surcriticalities $\xi=I_{dc}/I_{th}$ in turn lead to the lower amplitude relaxation rates. Indeed, the amplitude relaxation rate is defined as the balance between the damping and anti-damping rates \cite{slavin_2009-1, grimaldi_2014-1}. In the self-sustained regime, it is proportional to the surcriticality and can be expressed as $\Gamma_p=(\xi-1)\Gamma_G$ with $\Gamma_G$ the usual linear damping rate. Hence, close to the threshold for oscillations, the orbit’s relaxation takes longer time because of a smaller compensation of the magnetic damping. Consequently, a lower surcriticality induces a decrease of the amplitude relaxation rate $\Gamma_p$. 

Such lower amplitude relaxation rates $\Gamma_p$ at harmonic crossing indicate that the energy transfer between the two coupled modes is leading to the generation of an extra-damping term $\alpha_{coupling}$, affecting the dynamics of the excited mode \cite{romera_2015}. Given that the linear damping $\Gamma_G$ and the surcriticality $\xi$ are respectively proportional and inversely proportional to the damping constant $\alpha$ \cite{slavin_2009-1}, the amplitude relaxation rate $\Gamma_p$ is proportional to $1 - \alpha^2$. As a consequence, an enhancement of the damping constant leads to a decrease of the amplitude relaxation rate $\Gamma_p$ in agreement with our experimental results. The presence of a damped magnetic mode and the exchange of energy between hybridized magnetic modes can thus strongly modify the nonlinear behavior of the auto-oscillating mode excited by STT, allowing a fine tuning of the STO parameters.

\section{Self-resonant excitation driven by mode energy difference}

\begin{figure}
\includegraphics[scale=.2]{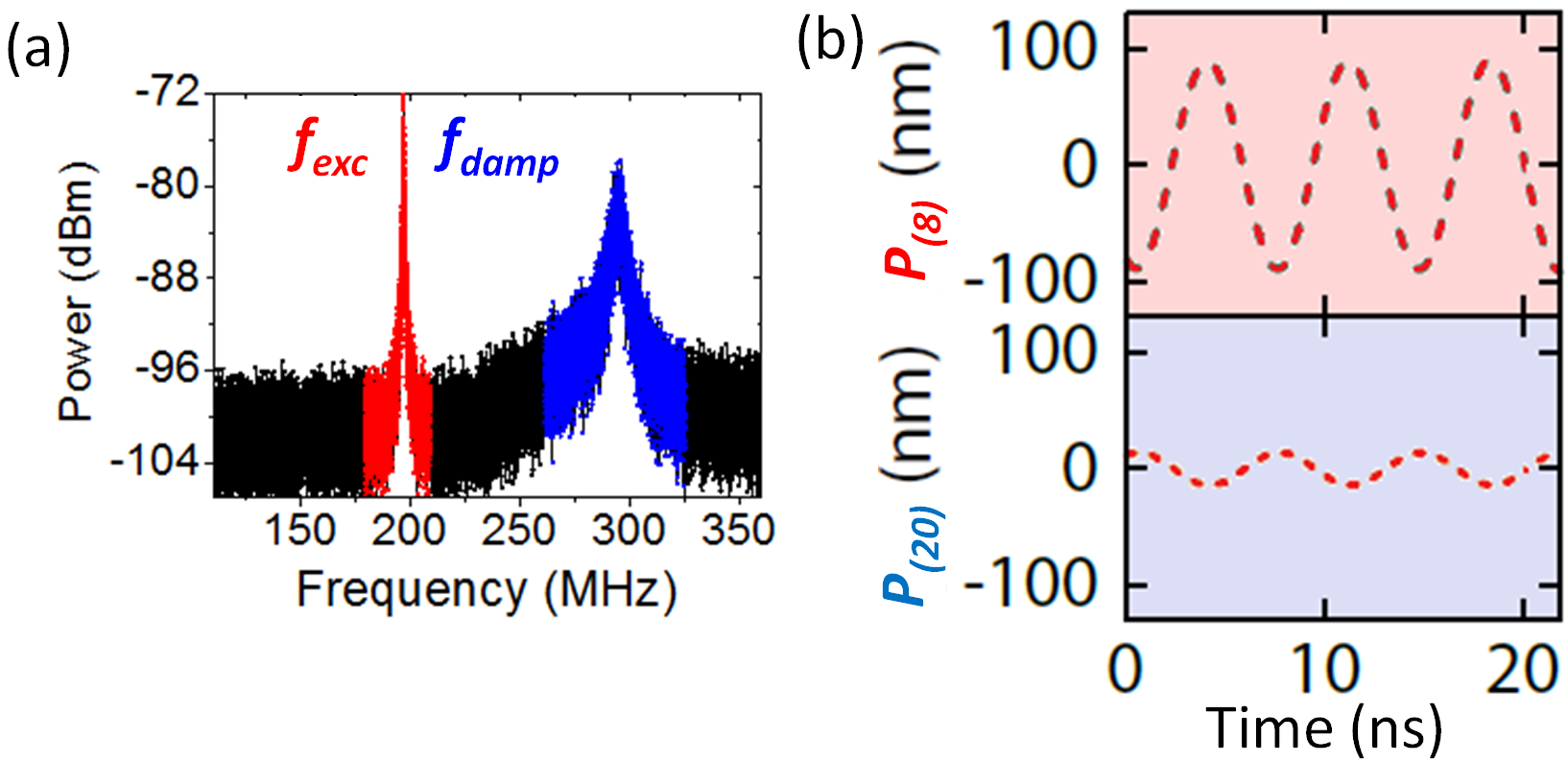}
\caption{(a) Frequency spectrum of the output emitted signal (associated with the dynamics in the bottom thick vortex layer close to the MgO barrier) at zero field for $I_{dc}=-16$ mA. The coupled mode mainly dominated by the thin layer is excited by STT while the coupled mode mainly dominated by the thick layer is damped by STT and only thermally activated (b) $\mu$Max micromagnetic simulations representing gyration radii of the mode excited by STT in the two vortex layers at zero field, $I_{dc} = -16$ mA. The simulations are performed at 0 K. The absence of thermal fluctuations explains that only the mode excited by STT is observed.}
\label{fig3}
\end{figure}

In order to simultaneously record the spectra corresponding to both the excited and damped modes, we have measured the properties of STOs from a second series in which the position of the thin and thick vortex layers in the stacking are inverted \footnote{To optimize the radio-frequency features, the thin vortex layer of these samples is of 8 nm instead of 6 nm.}. These samples are also NiFe based double vortex spin-valve above a magnetic tunnel junction with a SAF (NiFe(8nm)/Cu(9nm)/NiFe(20nm)/MgO/SAF). Consequently, both the excited and damped magnetic modes of the previous section can be probed through the detection of the magnetization dynamics of the bottom thick vortex layer which is close to the MgO barrier. These modes are here recorded for a negative current (and not positive as in the previous section) given that the positions of the two vortex layers have been exchanged. As shown in Fig. \ref{fig3}.a, we are able to simultaneously detect two peaks at zero field (see Fig. \ref{fig3}.a). The peak corresponding to the coupled mode excited by STT (in red) has a narrow linewidth of about 250 kHz and a frequency ($f_{exc}$) of 190 MHz. This mode, as confirmed by the micromagnetic simulations shown in Fig. \ref{fig3}.b, is mainly localized in the thin vortex layer excited by STT. One should note that this coupled mode is also weakly delocalized in the bottom thick vortex layer, close to the MgO barrier, allowing the detection of its dynamics. Indeed, the micromagnetic simulations show that the gyration radius of this mode is about 90 nm in the thin layer and around 10 nm in the bottom vortex layer. The second peak (in blue) observed experimentally in Fig. \ref{fig3}.a is much broader (linewidth above 4 MHz) and has a higher frequency ($f_{damp}$).  We attribute this peak to the thermal excitation of the second coupled mode (which explains its absence in the micromagnetic simulations that are performed at 0K).  It is to be noticed that the excited mode, mainly dominated by the thin vortex layer, has a lower frequency given that the vortex eigenfrequencies are proportional to the ratio $L/R$ with $L$ the thickness and $R$ the radius of the vortex layer \cite{guslienko_2002}.  
Surprisingly, the power of these peaks are experimentally of the same order. Given that the excited mode is only weakly delocalized in the bottom thick layer as observed in the simulations (90 nm vs 10 nm, see Fig. \ref{fig3}.b), its associated emitted power is of a similar order of magnitude as the one of the mode damped by STT and thermally activated. 

\begin{figure}
\centering
\includegraphics[scale=.3]{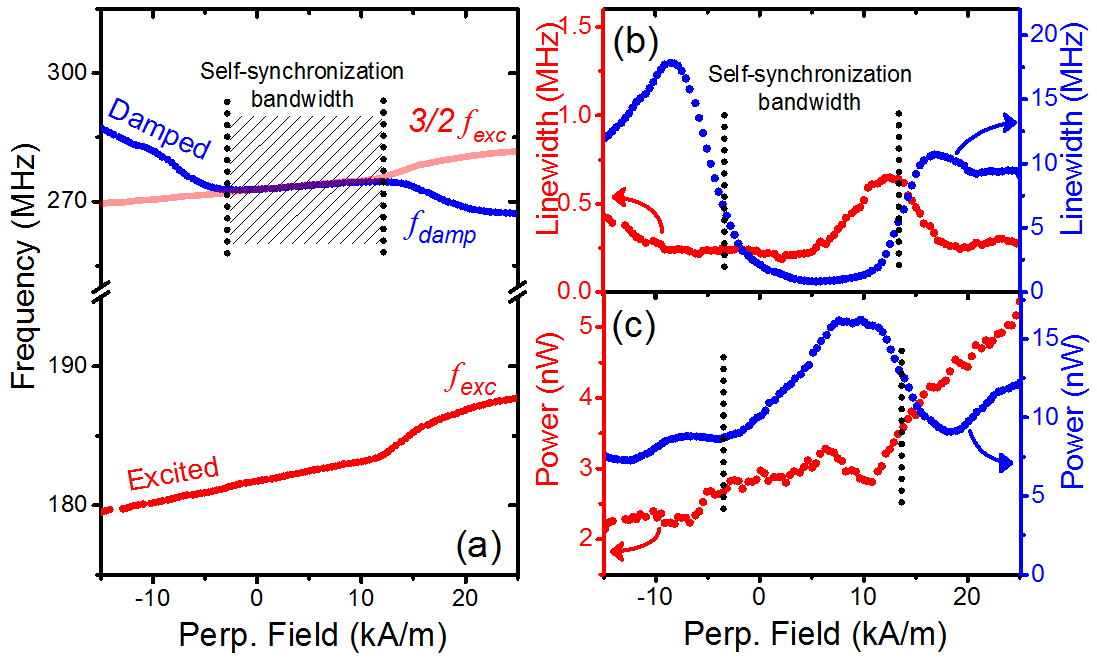}
\caption{Evolution of the frequencies (a), linewidths (b), powers (c) of both the excited (red dot) and damped (blue dots) coupled modes as a function of the applied perpendicular field $H_{perp}$. $I_{dc}$ is kept fixed at $-15$ mA.}
\label{fig:exchange_of_energy}
\end{figure}

In this section, we aim at identifying more precisely the mechanisms that drive the coupling between the two modes. In Fig. \ref{fig:exchange_of_energy}.a, we show the frequency evolution of each coupled mode as a function of the applied perpendicular field $H_{perp}$ for $I_{dc}= -15$ mA. For antiparallel cores configuration, the slopes of the modes are, as expected, of opposite sign \cite{hamadeh_2014-1, deloubens_2009, lebrun_2014} with the noticeable exception of the field region for which $f_{damp}=3f_{exc}/2$. In this crossing region (labelled as self-synchronization bandwidth in Fig. \ref{fig:exchange_of_energy}), the dynamics of the mode damped by STT appears to be enslaved by the auto-oscillating mode that is excited by STT. This results into a change of sign of the frequency evolution of the damped mode with $H_{perp}$ that becomes positive. In this field range, the damped mode becomes resonantly locked to the excited mode and its frequency is therefore equal to $3f_{exc}/2$. Moreover, the spectral linewidth of the damped mode is strongly decreased when $f_{damp}=3f_{exc}/2$ as seen in Fig. \ref{fig:exchange_of_energy}.b. At $H_{perp}= 5$ kA/m, its linewidth reaches a minimum of 800 kHz, close to the value of 350 kHz for the excited mode. We also notice that the linewidth of the damped mode is strongly enhanced at the edges of the locking region, which indicates the presence of phase slips, i.e. locking/unlocking events. These different features highlight that the dynamics of the damped mode is ``de facto'' resonantly driven by the other coupled mode excited by STT, hence the mechanism called ``self-resonance'' within the STO. 

Another important observation is that the interaction between the two modes is not unidirectional but bi-directional. Indeed we find that the linewidth of the excited mode increases in the field region corresponding to one of the two extremities of the self-resonance region. At these field values (around 10-15 kA/m), the frequency dispersion ($df_{exc}/dH_{perp}$) of the excited mode is enhanced. The mode interaction and the transfer of energy from the auto-oscillating mode to the damped mode are also illustrated by the evolution of their respective emitted powers. In the self-resonant region, the emitted power of the damped mode is enhanced from 10 to 15 nW while the detected power associated with the auto-oscillating mode decreases from 3 to 2.5 nW. One should notice that this correlation is only qualitative and not quantitative, given that, as we explained earlier, we probe only indirectly the dynamics of the auto-oscillating mode through the dynamics of the bottom thick vortex layer while most of its “actual” power is located in the top thin layer (see \ref{fig3}.b and Ref. \cite{lebrun_2014} for more details).

\begin{figure}
\centering
\includegraphics[scale=0.57]{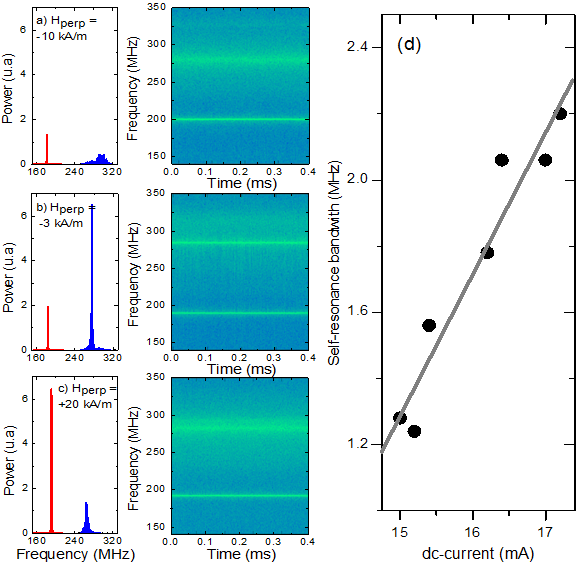}
\caption{Experimental power spectrum and instantaneous frequency time traces for three values of field: out of ($-10$ kA/m (a) and $+ 20$ kA/m (c)) and within ($+3$ kA/m (b)) the self-resonance region. $I_{dc}$ is kept fixed to $-15$ mA. (d) Evolution of the self-synchronization bandwidth as a function of the dc-current.}
\label{fig:self_reso_band}
\end{figure}

Recently, Iacocca et al. \cite{iacocca_2014} proposed to distinguish two regimes for describing the coupling between modes of a STO: mode hopping or mode coexistence. Thanks to real time measurements, we can directly demonstrate that our coupled dynamics correspond to a regime of mode coexistence. As shown in Fig. \ref{fig:self_reso_band}.a-c, the two modes coexist in time without any jumps between them in the whole range. In our study, contrary to Ref \cite{iacocca_2015}, the linewidths of the two coupled modes do not increase with the coupling strength (maximum at harmonic crossings) even though a mode co-existence is detected. It comes from the fact that the two interacting modes have the specificity of not being equally excited by STT. One mode is damped by STT and its dynamics mainly enslaved by the second one. The origin of the coupling mechanism that drives here the interaction between the two modes can either be mediated through spin-transfer and/or through dipolar \cite{pigeau_2012, hamadeh_2014-1, locatelli_2011} interaction. In Fig. \ref{fig:self_reso_band}.d, we show that the range of self-synchronization, determined by the derivative of the frequency of the damped mode, increases linearly as a function of the applied dc-current. If the coupling process would have been driven by STT, one should expect, for this current sign, an increase of the damping of the damped mode and thus a decrease of the self-synchronization bandwidth. On the contrary, the gyration motion associated with the dynamics of the excited mode in the top thin vortex layer increases with $I_{dc}$ for the experimental current sign. This larger gyration motion generates a larger dipolar field and thus a larger self-synchronization bandwidth. The experimental behavior thus indicates that the dynamics of the damped mode is in consequence driven by the amplitude of the dipolar field. Reciprocally, the forced dynamics of the damped mode in the thick layer influences the excited mode in the thin layer through dipolar coupling. There, both the effects of a non-static vortex polarizer on the STT and the presence of the rf-field emitted by the bottom vortex layer can destabilize the dynamics of the thin layer and explains the lower amplitude relaxation rate (compared to the case without coupling). These results shed light on the importance of mode coupling in STOs.

\section{Conclusion}

In summary, we have experimentally investigated the effects of mode coupling on the dynamics of a STO. By studying a double vortex based oscillators, we find clear evidences of the strong interaction between the coupled gyrotropic mode excited by STT and the other damped mode. The transfer of energy from one mode to the other one strongly increases the damping of the excited mode, reducing its power and its spectral coherence. When the coupling strength is maximum (for magnetic field corresponding to a mode crossing), we even report a forced excitation of the mode that is damped by STT through enslavement to the dynamics of the mode excited by STT. Beyond the understanding of the physic mechanisms that drive mode coupling, this self-resonant phenomena, here mediated by the internal rf-field of the oscillator, demonstrates the potential of mode coupling to develop multiband STOs.

The authors acknowledges M. Romera for fruitful discussions, Y. Nagamine, H. Maehara, and K. Tsunekawa of CANON ANELVA for preparing the MTJ films and the financial support from ANR agency (SPINNOVA ANR-11-NANO-0016) and EU FP7 grant (MOSAIC No. ICT-FP7- n.317950).

\email{V. Cros}
\bibliography{Bibliography}

\end{document}